\documentclass[doublespacing]{elsart}
\usepackage[english]{babel}

\usepackage{epsfig}
\usepackage{graphics}

\usepackage{amssymb}
\begin{document}

\begin{frontmatter}

\title{Depolarization of high-energy neutral particles in crystals and the possibility  to measure anomalous magnetic  moments of short-lived hyperons}

\author{V.G. Baryshevsky}

\address{Research Institute for Nuclear Problems, Belarusian State University, 11 Bobruiskaya str., 220030, Minsk, Belarus}

\begin{abstract}
The degree of  depolarization of neutral particles in crystals can reach tens of percents over the crystal length of  several centimeters,  which  can be the basis for possible experimental application of the depolarization effect for measuring anomalous magnetic moments of short-lived neutral hyperons.
\end{abstract}

\end{frontmatter}

\section*{Introduction}
\label{}

The increase in the proton beam energy up to 7 TeV, available at
the LHC, or  more, possible at the FCC, results in a growing
number of secondary particles available with these colliders.
According to \cite{1a,2a}, as the number of produced secondary
hyperons increases, we get a realistic hope to successfully use
the effects of charged hyperon spin rotation in bent crystals and
spin depolarization of high-energy particles moving in crystals
for measuring the anomalous magnetic moments of short-lived
hyperons \cite{1}.
Amongst hyperons, there is a family of neutral hyperons. Let us
note here that $\Xi^0_c$ hyperons have a lifetime $\tau\approx
1.1\cdot 1^{-13}$ s, $c\tau\approx 33.6 \mu$m ($c$ is the speed of
light), $\Omega^0_c$-hyperons -- $\tau\approx 7\cdot10^{-14}$ s,
$c\tau\approx 21 \mu$m, and $\Lambda^0_b$-hyperons -- $\tau\approx
10^{-12}$ s, $c\tau\approx 427\mu$m.

In this paper we demonstrate that spin depolarization of high-energy neutral hyperons moving in crystals at small angles to the crystal axes
(planes) can amount to tens of percent, allowing us to use the spin depolarization effect for measuring the magnetic moment of short-lived neutral hyperons. The estimated value of the running time for heavy neutral bottom baryons in FCC conditions is $\leq 300$ hours.

\section{Spin depolarization effect for neutral particles passing through crystals}

Let a particle move through a nonmagnetic crystal. The electric field $\vec{E}$ in the crystal takes on random values at the particle location point. As a result of fluctuations in the direction and magnitude of the field $\vec{E}$, the spin rotation angle of any transmitted particle  takes on different values, which leads to depolarization of  particle spin \cite{1}.

In considering depolarization of high-energy particles, we should bear in mind that their interaction with nuclei (atoms) of matter involves electromagnetic  as well as strong nuclear interactions.  From the quantum-mechanical viewpoint, the amplitude of particle scattering by a  scattering center, along with spin-independent terms also includes the terms depending on the spins of the incident particle and  the nuclei. We shall further be interested in the term proportional to the product $\vec S\vec n$, where $\vec S $ is the operator of the incident particle spin and $\vec n$ is the unit vector perpendicular to the particle scattering plane $\vec n=\frac{[\vec{p_0}\times \vec{p_1}]}{|\vec{p_0}\times\vec{p_1}|}$, where $\vec{p_0}$ is the initial particle momentum and $\vec{p_1}$ is the after-scattering particle momentum. This term is  called the Schwinger scattering in the case  when neutral particles are scattered in the nuclear electrostatic field. To describe the process of depolarization in the general case, we need to use the quantum kinetic equation for spin density matrix of a particle moving in the crystal \cite{3}. Nevertheless, the degree of depolarization can be estimated by considering the collisions between the particle and the crystal nuclei (atoms). To do this, we shall examine the particle-nucleus elastic scattering. We shall also assume that the nuclear spin is zero. In this case, the amplitude of elastic scattering of a spin $1/2$ hyperon by a nucleus can be written as  \cite{2}
\begin{equation}
\label{n1}
f= A+B\vec{\sigma} \vec{n},
\end{equation}
where $\vec{\sigma}= (\sigma_{x}, \sigma_{y},\sigma_z)$ are the Pauli spin matrices describing the particle spin $\vec S=\frac{1}{2}\vec\sigma$.
 The polarization vector of a particle that has undergone a single scattering event can be found using the following expression \cite{2,3}:
\begin{equation}
\label{n2}
\vec\xi= \frac{\mbox{tr}\rho f^+\vec\sigma f}{\mbox{tr}\rho f^+ f} =
\frac{\mbox{tr}\rho f^+ \sigma f}{\frac{d\sigma}{d\Omega}}.
\end{equation}
Here $\rho$ is the spin density matrix of the incident particle and
\begin{equation}
\label{n3}
\frac{d\sigma}{d\Omega}=\mbox{tr}\rho f^+ f = |A|^2 + |B|^2 + 2\texttt{Re} AB^*\vec n \vec\xi_0
\end{equation}
is the differential cross section of elastic scattering, where $\vec\xi_0$ is the beam's initial polarization $\vec\xi_0=\mbox{tr}\rho\vec\sigma$.
Equations (\ref{n1}) and (\ref{n2}) yield the following expression for the polarization vector of the scattered particle \cite{2}
\begin{equation}
\label{n4}
\vec\xi=\left\{(|A|^2 - |B|^2) \vec\xi_0 + 2 |B|^2 \vec n (\vec n\vec\xi_0)+ 2 \texttt{Im} (AB^*)[\vec n \vec\xi_0]
+2 \vec n \texttt{Re} (AB^*)\right\}\cdot \left(\frac{d\sigma}{d\Omega}\right)^{-1}.
\end{equation}
The contribution coming to the beam polarization $\vec\xi_N$ from particles scattered at an angle $\vartheta$ and having  the polarization vector $\vec\xi$ can be written in the form
\begin{equation}
\label{n4a}
\vec{\xi_N}{(\vartheta)}= \vec\xi\frac{d\sigma}{d\Omega}N^T_{nuc}j_0= \vec\xi\frac{d\sigma}{d\Omega}N_{nuc} L N_0,
\end{equation}
where $N_{nuc}$ is the density of target nuclei, $j_0$ is the current of incident $L$ is the target thickness, $N_0$ is the number of incident particles that have reached the target.
The total polarization of the scattered beam made up by summation of the contributions coming from polarization of every single particle is
\begin{equation}
\label{n5}
\vec\xi_N=\int\vec\xi (\Omega)\frac{d\sigma}{d\Omega}d\Omega N_{nuc}L N_0,
\end{equation}
i.e.,
\begin{eqnarray}
\label{n6}
\vec\xi_N  &=&  \int\left\{(|A|^2 - |B|^2) \vec\xi_0  + 2 |B|^2 \vec n (\vec n\vec\xi_0)+ 2 \texttt{Im} (AB^*)[\vec n \vec\xi_0]
+2 \vec n \texttt{Re} (AB^*)\right\}\nonumber\\
 &\times &d\Omega N_{nuc} L N_0.
\end{eqnarray}

Let us note that vector $\vec n$ lies in the  plane orthogonal to the direction of the initial momentum $\vec p_0$. That is why
integration over the  components of the scattered particle vector $\vec{p_1}$ that are transverse relative to vector $\vec{p_0}$ is
tantamount to integration over the direction of vector $\vec n$. After the integration, the last two terms in (\ref{n6}) vanish. Now, let us find how the beam polarization vector change after the beam has passed through the target. In the range of high energies, elastically scattered particles move at small angle, so we can first consider the situation when all particles (particles that passed through the target without collisions and particles transmitted
with scattering) reach the detector.

In this case, the number of particles entering the detector, $N_D$, is the same as $N_0$, the number of particles incident onto the target, and $N_D=N_0=N'_0+N_{sc}$, where $N'_0$ is the number of particles that have undergone collisionless transmission and $N_{sc}$ is the number of scattered particles.
Hence, the beam polarization change $\Delta\vec\xi_N$ is
\begin{eqnarray}
\label{n7}
\Delta\vec\xi_N=\vec\xi_N - \vec\xi_{0N} & = & \int\left\{(|A|^2 - |B|^2) \vec\xi_0  + 2 |B|^2 \vec n (\vec n\vec\xi_0)+ 2 \texttt{Im} (AB^*)
[\vec n \vec\xi_0]+2 \vec n \texttt{Re} (AB^*)\right\}  \nonumber\\
&\times & N_{nuc} L N_0- \vec \xi_0\int\frac{d\sigma}{d\Omega} d\Omega N_{nuc} L N_0.
\end{eqnarray}
Now let us find the degree of longitudinal depolarization. We shall assume that the initial polarization vector $\vec\xi_0$ is directed along the particle momentum; as a consequence $\vec n \perp \vec \xi_0$. In this case, the value of longitudinal depolarization
\begin{equation}
\label{n8}
\eta_{\parallel}=\frac{\xi_{N\parallel}-\xi_{0N}}{\xi_{0N}}= -2\int|B|^2 d\Omega N_{nuc} L.
\end{equation}
We can verify that $\eta_{\perp}= \frac{1}{2}\eta_{\parallel}$. Thus, the value of depolarization is determined by the amplitude $B$ of spin-orbit scattering (the spin-orbit scattering cross section $\int|B|^2 d\Omega$).

In the case considered here, the amplitude $B$ is determined by
two contributions: the nuclear interaction of the neutral particle
(hyperon or neutron) and the interaction between the anomalous
magnetic moment and the nuclear (atomic) electrostatic field, also
known as the Schwinger interaction \cite{4}. Depolarization of
neutral particles due to Schwinger interaction  in amorphous media
was discussed by V. Lyuboshitz \cite{5}. According to \cite{5},
the expressions for the degree of depolarization in this case can
be written in the form:
\begin{equation}
\label{n9}
\eta_{\parallel}\simeq \frac{1}{8}\left(\frac{E_s}{m_p}\right)^2 g^2\frac{L}{L_{rad}},
\end{equation}
where $E_s= 21$ MeV, $m_p$ is the proton mass, $g$ is the particle
$g$-factor, and $L_{rad}$ is the radiation length. According to
(\ref{n9}), $\eta_{\parallel}$ is independent of the particle
energy. For neutrons, the depolarization degree $\eta_{\parallel}
\simeq 10^{-3}\frac{L}{L_{rad}}$, whereas for $\Lambda$-hyperons
it is $10^{-4}\frac{L}{L_{rad}}$ and
$\eta_{\perp}=\frac{1}{2}\eta_{\parallel}$. As is seen, in
amorphous medium, the depolarization degree due to Schwinger
scattering at a length of several centimeters is even less than
$\sim 10^{-2}$. In the range of high energies, nuclear interaction
should also be considered together with the Schwinger one. A
substantially greater degree of depolarization occurs when neutral
particles pass through crystals at small angles to
crystallographic axes (planes). In this case, the cross section
for scattering by a single nucleus  can rise ten- or even
hundred-fold. In particular, it was shown in \cite{6,7} that for
neutrons with energies up to several MeV, the cross section for
coherent elastic Schwinger scattering by a single nucleus in
crystals increases  by a factor of ten. According to \cite{7}, the
total cross section for Schwinger scattering of 1 MeV neutrons in
crystals  undergoes oscillations, depending on the particle
entering angle, reaching in tungsten crystals the values
$\sim2\cdot 10^{-25}$ cm$^2$ , which is 10 times as high as that
for Schwinger scattering  by a nucleus in amorphous media:
$\sigma= 1.6\cdot 10 ^{-26}$ cm$^2$.

The scattering cross section of particles in crystals can be analyzed using the following expression \cite{3}:
\begin{equation}
\label{n10}
\frac{d \sigma_{cr}}{d\Omega} = \frac{d\sigma}{d\Omega_+}\frac{1}{N}\overline{\left|\sum^N_{n=1}\exp(i\vec q\vec{r_n})\right|^2}.
\end{equation}
Here $\vec{r_n}$ is the radius vector of the $n$-th nucleus in the crystal, $\vec q= \vec k'-\vec k$ is the transmitted momentum, $N$ is the number of scattering centers, $\vec{k’}= \frac{\vec {p_1}}{\hbar} $, and $\vec k=\vec{p_0}{\hbar}$. The overline means averaging over thermal oscillations of the lattice nuclei (atoms); $\frac{d\sigma}{d\Omega} = \texttt{tr}\rho f^+(\vec q)\hat{f}(\vec q) $, where $f(\vec q) $ is the amplitude of coherent elastic scattering of  the spin $\vec S$ particle by the nucleus. In the general case,  if the nuclei  are polarized, $f$ depends on their polarization (for detail, see \cite{3}).

Averaging over thermal oscillations can be performed in the same manner as it is done in the theory of X-ray scattering \cite{9,10}

As a result, we have
\begin{equation}
\label{n11}
\frac{d\sigma_{cr}}{d\Omega}=\frac{d\sigma}{d\Omega}\left\{(1-e^{-\overline{u^2}
{q^2}}) +\frac{1}{N}\left|\sum_n e^{i\vec q\vec{r}_n^0}\right|^2 e^{-\overline{u^2} {q^2}}\right\},
\end{equation}
where $\vec{r}_n^0 $ is the coordinate of the center of gravity of the crystal  nucleus, $\overline{u^2}$ is the mean square  of thermal oscillations of nuclei in the crystal. The first term  describes incoherent scattering and the second one describes the coherent coherent due to periodic arrangement of crystal nuclei (atoms). This contribution leads to the increase in the cross section.  This expression can be used as long as the  crystal length satisfies the  inequality $k(n-1) L\ll 1$, where $n$ is the particle refractive index in the crystal \cite{3}.

  The depolarization degree $\eta_{\parallel}^{cr}$ in crystals can be estimated  using the following  expression for the degree of longitudinal
  depolarization:

\[
\eta_{\parallel}^{cr}(\vartheta_0)= \frac{\sigma_{cr}(\vartheta_0)}{\sigma}\eta_{\parallel},
\]

where  $\sigma_{cr}$ is the part of the total scattering cross
section in the crystal due to spin-orbit (Schwinger scattering)
interaction, $\sigma$ is the part of the total scattering cross
section by the nucleus (atom) in the amorphous medium that is  due
to  spin-orbit interaction (Schwinger interaction), $\vartheta_0$
is the angle of particle entrance relative to the crystallographic
axis (plane), and $\eta_{\parallel}$ is the degree of particle
depolarization in the amorphous medium (see equation(\ref{n9})).

The degree of depolarization $\eta_{\parallel}^{cr}(\vartheta_0)$ (and $\eta_{\perp}^{cr}=\frac{1}{2}\eta_{\parallel}^{cr}(\vartheta_0)$ depends on the
angle of particle entrance into the crystal relative to the axis (plane) and undergoes oscillations  similar to those observed in coherent bremsstrahlung
and pair production, e.g., in the crystal (see, e.g. \cite{10}).  According to \cite{3}, the increase in the cross section also refers to particle scattering in crystals with polarized
nuclei.                                                                                                                                                      

As the particle energy increases, the coherent elastic scattering cross section in crystals also increases, because the coherent length grows. As a
consequence, the degree of particle depolarization also increases, reaching tens of percents over the tungsten crystal length of several centimeters.
The increase in the particle energy is accompanied by the growing number of produced hyperons. As a
result, the measurements of the spin depolarization value of short-lived neutral hyperons in crystals gives hope for measuring their magnetic moment at the
LHC and the FCC.

\section{Conclusion}
The degree of neutral particle (hyperon, neutron)  depolarization in crystals increases as the particle energy is increased, because the scattering coherent length grows.

 The degree of  depolarization of neutral particles in crystals can reach tens of percents over the crystal
 length of  several centimeters,  which  can be the basis for possible experimental application of the depolarization effect \cite{1}
 for measuring anomalous magnetic moments of short-lived neutral hyperons. Depolarization of neutral $\Lambda$-hyperons can be used for calibration of the experiment, because their magnetic moment is known. They have
a longer lifetime ($\tau=2.6\cdot 10^{-10}$~s) and so other
techniques apply to measure their anomalous magnetic moment. The
estimates show  that the time $T$ for observation  of this effect
for neutral charm hyperons  is of  the same order as that of spin
rotation effect for charged $\Lambda^+_c$-hyperons in a bent
crystal reported in \cite{1a} $T\approx10\div 20$ hours.  The
production rate for neutral bottom baryons is noticeably less, but
increases as the proton energy is increased, giving us hope  that
the experiments on measuring their anomalous magnetic moments
during tens of hours (up to 300 hours) are possible.

\end{document}